\documentclass[12pt]{article}
\usepackage{amsmath,amsfonts,amssymb}  

\makeatletter
\def\numberbysection{\@addtoreset{equation}{section}
    \def\theequation{\thesection.\arabic{equation}}}
\makeatother

\numberbysection

\newcommand{\be}{\begin{eqnarray}}
\newcommand{\ee}{\end{eqnarray}}
\newcommand{\non}{\nonumber}
\newcommand{\id}{\mathbb{I}}

\newcommand{\sgn}{\mathop{\rm sign}\nolimits}
\newcommand{\diag}{\mathop{\rm diag}\nolimits}
\newcommand{\cf}{\ensuremath{\mathfrak{c}}}

\begin{document}

\begin{titlepage}
\strut\hfill UMTG--256
\vspace{.5in}
\begin{center}

\LARGE The Zamolodchikov-Faddeev algebra\\
for open strings attached to giant gravitons\\
\vspace{1in}
\large Changrim Ahn \footnote{
       Department of Physics, Ewha Womans University, 
       Seoul 120-750, South Korea} and
       Rafael I. Nepomechie \footnote{
       Physics Department, P.O. Box 248046, University of Miami,
       Coral Gables, FL 33124 USA}\\

\end{center}

\vspace{.5in}

\begin{abstract}
We extend the Zamolodchikov-Faddeev algebra for the superstring sigma
model on $AdS_{5}\times S^{5}$, which was formulated by Arutyunov,
Frolov and Zamaklar, to the case of open strings attached to maximal
giant gravitons, which was recently considered by Hofman and
Maldacena.  We obtain boundary $S$-matrices which satisfy the standard
boundary Yang-Baxter equation.
\end{abstract}

\end{titlepage}

\setcounter{footnote}{0}

\section{Introduction}\label{sec:intro}

A factorizable $S$-matrix \cite{ZZ, Fa} describing the scattering of
world-sheet excitations of the $AdS_{5}\times S^{5}$ superstring sigma
model \cite{MT} has been proposed by Arutyunov, Frolov and Zamaklar  (AFZ)
\cite{AFZ}.  This $S$-matrix is closely related to the one found 
earlier by Beisert \cite{Be} describing the scattering of excitations of the
dynamic spin chain corresponding to planar ${\cal N}=4$ super
Yang-Mills.  However, the AFZ ``string'' $S$-matrix obeys the
standard Yang-Baxter equation, while Beisert's $S$-matrix obeys a twisted
(dynamical) Yang-Baxter equation. \footnote{There are in fact three 
relevant $S$-matrices: $S^{string}_{AFZ}$, which is in the ``string'' 
basis, and satisfies the standard YBE; $S^{chain}_{AFZ}$, which is 
in the ``spin chain'' basis, and satisfies a twisted YBE; and 
$S_{Beisert}$, which is related to $S^{chain}_{AFZ}$ by the final 
(unnumbered) equation of Section 7 in \cite{AFZ}.}
The string $S$-matrix (up to a
phase) follows directly from the assumption that the excitations are
described by a Zamolodchikov-Faddeev (ZF) algebra, and that they have a
centrally extended $su(2|2) \oplus su(2|2)$ symmetry \cite{Be,
AFPZ}.  It agrees with perturbative results obtained by direct
computations \cite{KMRZ}.

Hofman and Maldacena (HM) \cite{HM} recently considered open strings
attached to maximal giant gravitons \cite{giants} in $AdS_{5}\times
S^{5}$.  (Related earlier work includes \cite{BV, Ag, MV}.)  They
proposed boundary $S$-matrices describing the reflection of
world-sheet excitations (giant magnons) for two cases, namely, the
$Y=0$ and $Z=0$ giant graviton branes.  However, we have found that
the boundary $S$-matrix for the latter case does not satisfy the
standard boundary Yang- Baxter equation (BYBE) \cite{Ch, GZ}.

The purpose of this note is to construct related boundary $S$-matrices
which do obey the standard BYBE. To this end, we extend the ZF algebra
which was formulated by AFZ by introducing boundary operators with
suitable symmetry properties.  We explicitly verify that the resulting
boundary $S$-matrices are indeed solutions of the standard BYBE.

The outline of this paper is as follows.  In Section \ref{sec:bulk} we
briefly review the bulk ZF algebra and the computation of the bulk
$S$-matrix, which in fact is the transpose of the matrix given in
\cite{AFZ}.  In Section \ref{sec:boundary} we formulate the boundary
ZF algebra, and present boundary $S$-matrices for both the $Y=0$ and
$Z=0$ giant graviton branes.  In Section \ref{sec:crossing} we derive
crossing relations for the boundary $S$-matrices and solve for the
corresponding scalar factors.  We conclude in Section
\ref{sec:conclude} with a brief discussion of our results.

\section{Bulk ZF algebra and $S$-matrix}\label{sec:bulk}

In this section, we briefly review the bulk ZF algebra and the
computation of the bulk $S$-matrix.  \footnote{We consider the
$S$-matrix corresponding to a single copy of the centrally extended
$su(2|2)$ algebra; the full $S$-matrix is a tensor product of two such
$S$-matrices.}
Following AFZ \cite{AFZ}, we
denote the ZF operators by $A_{i}^{\dagger}(p)$, $i=1\,, 2\,, 3\,, 4$.
These operators create asymptotic particle states of momentum $p$ when
acting on the vacuum state $|0\rangle$.  The bulk $S$-matrix elements
$S_{i\, j}^{i' j'}(p_{1}, p_{2})$ are defined by the relation
\be
A_{i}^{\dagger}(p_{1})\, A_{j}^{\dagger}(p_{2}) = 
S_{i\, j}^{i' j'}(p_{1}, p_{2})\, 
A_{j'}^{\dagger}(p_{2})\, A_{i'}^{\dagger}(p_{1}) \,,
\label{bulkS1}
\ee
where summation over repeated indices is always understood.  It is
convenient to arrange these matrix elements into a $16 \times 16$
matrix $S$ as follows,
\be
S = S_{i\, j}^{i' j'} e_{i\, i'}\otimes e_{j\, j'}\,,
\label{bulkS2}
\ee
where $e_{i j}$ is the usual elementary $4 \times 4$ matrix whose 
$(i, j)$ matrix element is 1, and all others are zero.  Although
(\ref{bulkS2}) is the standard convention, AFZ use a different
convention (see Eq.  (8.4) in \cite{AFZ}), such that our matrix $S$ is
the {\it transpose} of theirs.

As is well known \cite{ZZ}, starting from $A_{i}^{\dagger}(p_{1})\,
A_{j}^{\dagger}(p_{2})\, A_{k}^{\dagger}(p_{3})$, one can arrive at
linear combinations of $A_{k''}^{\dagger}(p_{3})\,
A_{j''}^{\dagger}(p_{2})\, A_{i''}^{\dagger}(p_{1})$ by applying the
relation (\ref{bulkS1}) three times, in two different ways.  The 
consistency condition is the Yang-Baxter equation,
\be
S_{12}(p_{1}, p_{2})\, S_{13}(p_{1}, p_{3})\, S_{23}(p_{2}, p_{3})\ =
S_{23}(p_{2}, p_{3})\, S_{13}(p_{1}, p_{3})\, S_{12}(p_{1}, p_{2}) \,.
\label{YBE}
\ee
We use the standard convention $S_{12} = S \otimes \id$, $S_{23} 
= \id \otimes S$, and $S_{13} = {\cal P}_{12}\, S_{23}\, {\cal P}_{12}$,
where ${\cal P}_{12} = {\cal P} \otimes \id$, ${\cal P} = e_{i\, j} \otimes
e_{j\, i}$ is the permutation matrix, and $\id$ is the four-dimensional
identity matrix. The ZF algebra (\ref{bulkS1}) also implies the bulk
unitarity equation
\be
S_{12}(p_{1}, p_{2})\, S_{21}(p_{2}, p_{1}) = \id \,,
\label{bulkunitarity}
\ee
where $S_{21} = {\cal P}_{12}\, S_{12}\, {\cal P}_{12}$.

For later reference, we note (as also discussed in \cite{AFZ}) that the
conjugate operators $\left( A_{i}^{\dagger}(p) \right)^{\dagger} =
A^{i}(p)$ obey
\be
A^{i}(p_{1})\, A^{j}(p_{2}) = 
S^{\, i\, j}_{i' j'}(p_{1}, p_{2})\, 
A^{j'}(p_{2})\, A^{i'}(p_{1}) \,,
\label{bulkSconjugate}
\ee
which together with (\ref{bulkS1}) implies the so-called physical unitarity 
condition $S_{21}(p_{2}, p_{1}) = S_{12}^{\dagger}(p_{1}, p_{2})$, 
and therefore
\be
S_{12}(p_{1}, p_{2})\, S_{12}^{\dagger}(p_{1}, p_{2}) = \id \,.
\label{bulkphysicalunitarity}
\ee 

The centrally extended $su(2|2)$ algebra consists of the rotation
generators $L_{a}^{\ b}$, $R_{\alpha}^{\ \beta}$, the supersymmetry
generators $Q_{\alpha}^{\ a}$, $Q_{a}^{\dagger \alpha}$, and the
central elements $C\,, C^{\dagger}\,, H$.  Latin indices $a\,, b\,, 
\ldots$ take values $\{1\,, 2\}$, while Greek indices $\alpha\,, 
\beta\,, \ldots$ take values $\{3\,, 4\}$. These generators have the
following nontrivial commutation relations \cite{AFZ, Be, IK}
\be
\left[ L_{a}^{\ b}\,, J_{c} \right] &=& \delta_{c}^{b} J_{a} - 
\frac{1}{2} \delta_{a}^{b} J_{c}\,, \quad 
\left[ R_{\alpha}^{\ \beta}\,, J_{\gamma} \right] =
\delta_{\gamma}^{\beta} J_{\alpha} - 
\frac{1}{2} \delta_{\alpha}^{\beta} J_{\gamma}\,, \non  \\
\left[ L_{a}^{\ b}\,, J^{c} \right] &=& -\delta_{a}^{c} J^{b} + 
\frac{1}{2} \delta_{a}^{b} J^{c}\,, \quad 
\left[ R_{\alpha}^{\ \beta}\,, J^{\gamma} \right] =
-\delta_{\alpha}^{\gamma} J^{\beta} +
\frac{1}{2} \delta_{\alpha}^{\beta} J^{\gamma}\,, \non \\
\Big\{Q_{\alpha}^{\ a}\,, Q_{\beta}^{\ b}\Big\}&=& 
\epsilon_{\alpha \beta}\epsilon^{a b} C \,, \quad 
\Big\{Q_{a}^{\dagger \alpha}\,, Q_{b}^{\dagger \beta} \Big\}=
\epsilon^{\alpha \beta}\epsilon_{a b} C^{\dagger} \,, \non \\
\Big\{Q_{\alpha}^{\ a}\,, Q_{b}^{\dagger \beta} \Big\} &=& \delta_{b}^{a} 
R_{\alpha}^{\ \beta}+ \delta_{\alpha}^{\beta} L_{b}^{\ a} 
+ \frac{1}{2} \delta_{b}^{a} \delta_{\alpha}^{\beta} H \,,
\label{symmetryalgebra}
\ee
where $J_{i}$ ($J^{i}$) denotes any lower (upper) index of a generator,
respectively. 

The action of the bosonic generators on the ZF operators is given by 
\be
L_{a}^{\ b}\, A_{c}^{\dagger}(p) &=& (\delta_{c}^{b}\delta_{a}^{d} - 
\frac{1}{2}\delta_{a}^{b}\delta_{c}^{d}) A_{d}^{\dagger}(p) +
A_{c}^{\dagger}(p)\, L_{a}^{\ b} \,, \quad 
L_{a}^{\ b}\, A_{\gamma}^{\dagger}(p) = A_{\gamma}^{\dagger}(p)\, 
L_{a}^{\ b}\,, \non \\
R_{\alpha}^{\ \beta}\, A_{\gamma}^{\dagger}(p) &=& 
(\delta_{\gamma}^{\beta}\delta_{\alpha}^{\delta} - 
\frac{1}{2}\delta_{\alpha}^{\beta}\delta_{\gamma}^{\delta}) 
A_{\delta}^{\dagger}(p) +
A_{\gamma}^{\dagger}(p)\, R_{\alpha}^{\ \beta} \,, \quad
R_{\alpha}^{\ \beta}\, A_{c}^{\dagger}(p) =  A_{c}^{\dagger}(p)\, 
R_{\alpha}^{\ \beta}\,.
\label{repBulk1}
\ee
Moreover, the action of the supersymmetry generators is given by (see
Eq.  (4.21) in \cite{AFZ})
\be
Q_{\alpha}^{\ a}\, A_{b}^{\dagger}(p) &=& e^{- i p/2} \left[
a(p) \delta_{b}^{a} A_{\alpha}^{\dagger}(p) + 
A_{b}^{\dagger}(p)\, Q_{\alpha}^{\ a} \right] \,, \non \\
Q_{\alpha}^{\ a}\, A_{\beta}^{\dagger}(p) &=& e^{- i p/2} \left[ 
b(p) \epsilon_{\alpha \beta}\epsilon^{a b} A_{b}^{\dagger}(p) -
A_{\beta}^{\dagger}(p)\, Q_{\alpha}^{\ a} \right]\,, \non \\
Q_{a}^{\dagger \alpha}\, A_{b}^{\dagger}(p) &=& e^{i p/2} \left[
c(p) \epsilon_{a b} \epsilon^{\alpha \beta} A_{\beta}^{\dagger}(p) + 
A_{b}^{\dagger}(p)\, Q_{a}^{\dagger \alpha} \right] \,, \non \\
Q_{a}^{\dagger \alpha}\, A_{\beta}^{\dagger}(p) &=& e^{i p/2} \left[
d(p) \delta_{\beta}^{\alpha} A_{a}^{\dagger}(p) - 
A_{\beta}^{\dagger}(p)\, Q_{a}^{\dagger \alpha} \right] \,.
\label{repBulk2}
\ee 
AFZ work with a different set of relations for the supersymmetry
generators which involve the world-sheet momentum operator (see Eq.
(4.15) in \cite{AFZ}).  However, as we shall see in Section
\ref{subsec:Z}, the relations (\ref{repBulk2}) are more natural
when dealing with a boundary.

The ZF operators form a representation of the symmetry algebra with $C
= a\, b\, e^{-i p}\,, \quad  C^{*} = c\, d\, e^{i p}\,, \ H= a d + b c$,
provided $a d - b c = 1$.  The representation is also unitary provided
$d= a^{*}\,, c = b^{*}$.  Since $C = i g (1-e^{-i p})$ \cite{AFZ},
the parameters can be chosen as follows \cite{AFZ, Be}
\be
a = \sqrt{g}\eta\,, \quad 
b = \sqrt{g}\frac{i}{\eta}\left(\frac{x^{+}}{x^{-}}-1\right)\,, \quad 
c= -\sqrt{g}\frac{\eta}{x^{+}}\,, \quad 
d=\sqrt{g}\frac{x^{+}}{i \eta}\left(1 - \frac{x^{-}}{x^{+}}\right)\,,
\label{BulkParameters}
\ee
where
\be
x^{+}+\frac{1}{x^{+}}-x^{-}-\frac{1}{x^{-}} = \frac{i}{g}\,, \quad 
\frac{x^{+}}{x^{-}} = e^{i p}\,, \quad 
\eta = \sqrt{i(x^{-}-x^{+})} \,.
\label{eta}
\ee 
Hence,
\be
H = -i g \left(x^{+}-\frac{1}{x^{+}}-x^{-}+\frac{1}{x^{-}}\right) \,.
\ee 

The $S$-matrix can be determined (up to a phase) by demanding that it
commute with the symmetry generators. \footnote{The idea of using 
nonlocal (fractional-spin) integrals of motion to determine bulk
$S$-matrices goes at least as far back as the works \cite{KR, Za}. 
This approach was extended to boundary $S$-matrices in \cite{boundQG}.} 
That is, starting from
$J\, A_{i}^{\dagger}(p_{1})\, A_{j}^{\dagger}(p_{2})
|0\rangle$ where $J$ is a symmetry generator, and assuming
that $J$ annihilates the vacuum state, one can
arrive at linear combinations of $A_{j'}^{\dagger}(p_{2})\,
A_{i'}^{\dagger}(p_{1}) |0\rangle$ in two different ways, by applying the
ZF relation (\ref{bulkS1}) and the symmetry relations 
(\ref{repBulk1}), (\ref{repBulk2})
in different orders.  The consistency condition is
a system of linear equations for the $S$-matrix
elements.  The result for the nonzero matrix elements is \cite{AFZ}
\be
S_{a\, a}^{a\, a} &=& A\,, \quad 
S_{\alpha\, \alpha}^{\alpha\, \alpha} = D\,, \non \\
S_{a\, b}^{a\, b} &=& \frac{1}{2}(A-B)\,, \quad 
S_{a\, b}^{b\, a} = \frac{1}{2}(A+B) \,, \non \\
S_{\alpha\, \beta}^{\alpha\, \beta} &=& \frac{1}{2}(D-E)\,, \quad 
S_{\alpha\, \beta}^{\beta\, \alpha} = \frac{1}{2}(D+E) \,, \non \\
S_{a\, b}^{\alpha\, \beta} &=& 
-\frac{1}{2}\epsilon_{a b}\epsilon^{\alpha \beta}\, C \,, \quad
S_{\alpha\, \beta}^{a\, b} = 
-\frac{1}{2}\epsilon^{a b}\epsilon_{\alpha \beta}\, F \,, \non \\
S_{a\, \alpha}^{a\, \alpha} &=& G\,, \quad 
S_{a\, \alpha}^{\alpha\, a} = H \,, \quad 
S_{\alpha\, a}^{a\, \alpha} = K\,, \quad 
S_{\alpha\, a}^{\alpha\, a} = L \,,  
\label{bulkS3}
\ee
where $a\,, b \in \{1\,, 2\}$ with $a \ne b$;  
$\alpha\,, \beta \in \{3\,, 4\}$ with $\alpha \ne \beta$; and
\be
A &=& S_{0}\frac{x^{-}_{2}-x^{+}_{1}}{x^{+}_{2}-x^{-}_{1}}
\frac{\eta_{1}\eta_{2}}{\tilde\eta_{1}\tilde\eta_{2}} \,, \non \\
B &=&-S_{0}\left[\frac{x^{-}_{2}-x^{+}_{1}}{x^{+}_{2}-x^{-}_{1}}+
2\frac{(x^{-}_{1}-x^{+}_{1})(x^{-}_{2}-x^{+}_{2})(x^{-}_{2}+x^{+}_{1})}
{(x^{-}_{1}-x^{+}_{2})(x^{-}_{1}x^{-}_{2}-x^{+}_{1}x^{+}_{2})}\right]
\frac{\eta_{1}\eta_{2}}{\tilde\eta_{1}\tilde\eta_{2}}\,, \non \\
C &=& S_{0}\frac{2i x^{-}_{1} x^{-}_{2}(x^{+}_{1}-x^{+}_{2}) \eta_{1} \eta_{2}}
{x^{+}_{1} x^{+}_{2}(x^{-}_{1}-x^{+}_{2})(1 - x^{-}_{1} x^{-}_{2})} 
\,, \qquad
D = -S_{0}\,, \non \\
E &=&S_{0}\left[1-2\frac{(x^{-}_{1}-x^{+}_{1})(x^{-}_{2}-x^{+}_{2})
(x^{-}_{1}+x^{+}_{2})}
{(x^{-}_{1}-x^{+}_{2})(x^{-}_{1} x^{-}_{2}-x^{+}_{1} 
x^{+}_{2})}\right]\,, \non \\
F &=& S_{0}\frac{2i(x^{-}_{1}-x^{+}_{1})(x^{-}_{2}-x^{+}_{2})(x^{+}_{1}-x^{+}_{2})}
{(x^{-}_{1}-x^{+}_{2})(1-x^{-}_{1} x^{-}_{2})\tilde\eta_{1} \tilde\eta_{2}}\,, 
\non \\
G &=&S_{0}\frac{(x^{-}_{2}-x^{-}_{1})}{(x^{+}_{2}-x^{-}_{1})}
\frac{\eta_{1}}{\tilde\eta_{1}}\,, \qquad 
H =S_{0}\frac{(x^{+}_{2}-x^{-}_{2})}{(x^{-}_{1}-x^{+}_{2})}
\frac{\eta_{1}}{\tilde\eta_{2}}\,, \non \\
K &=&S_{0}\frac{(x^{+}_{1}-x^{-}_{1})}{(x^{-}_{1}-x^{+}_{2})}
\frac{\eta_{2}}{\tilde\eta_{1}}\,, \qquad 
L = S_{0}\frac{(x^{+}_{1}-x^{+}_{2})}{(x^{-}_{1}-x^{+}_{2})}
\frac{\eta_{2}}{\tilde\eta_{2}}\,, 
\label{bulkS4}
\ee
where 
\be
x^{\pm}_{i} = x^{\pm}(p_{i})\,, \quad 
\eta_{1} = \eta(p_{1}) e^{i p_{2}/2}\,, \quad \eta_{2}=\eta(p_{2})\,, 
\quad \tilde\eta_{1} =\eta(p_{1})\,, \quad \tilde\eta_{2} 
=\eta(p_{2})e^{i p_{1}/2}\,,
\ee
and $\eta(p)$ is given in (\ref{eta}).  This $S$-matrix satisfies the
standard Yang-Baxter equation (\ref{YBE}). It also satisfies the 
unitarity equation (\ref{bulkunitarity}), provided that the scalar 
factor obey
\be
S_{0}(p_{1}, p_{2})\, S_{0}(p_{2}, p_{1}) = 1 \,.
\label{bulkunitarityscalar}
\ee

\section{Boundary ZF algebra and $S$-matrix}\label{sec:boundary}

We consider now the problem of scattering from a boundary. Following 
HM \cite{HM}, we consider the cases of the $Y=0$ and $Z=0$ giant 
graviton branes, which we consider in turn.

\subsection{$Y=0$ giant graviton brane}\label{subsec:Y}

In order to describe boundary scattering, we extend the bulk ZF
algebra (\ref{bulkS1}) by introducing appropriate boundary operators
which create the boundary-theory vacuum state $|0\rangle_{B}$ from
$|0\rangle$ \cite{GZ}.  Since there is no boundary degree of freedom
for the $Y=0$ giant graviton brane, the corresponding boundary
operator is a scalar.  For a right boundary, we introduce a right
boundary operator $B_{R}$, and define the right boundary $S$-matrix by
\be
A^{\dagger}_{i}(p)\, B_{R} = R_{\, i}^{R\, i'}(p)\, A^{\dagger}_{i'}(-p)\,  
B_{R} \,.
\label{boundarySRY}
\ee
We arrange the $S$-matrix elements in the usual way into a matrix 
$R^{R} = R_{\, i}^{R\, i'}\, e_{i\, i'}$.
Starting from $A^{\dagger}_{i}(p_{1})\, A^{\dagger}_{j}(p_{2})\, B_{R}$, 
one can arrive at linear combinations of 
$A^{\dagger}_{i'''}(-p_{1})\, A^{\dagger}_{j'''}(-p_{2})\, B_{R}$
by applying each of the relations (\ref{bulkS1}) and 
(\ref{boundarySRY}) two times, in two different ways. The consistency 
condition is the right BYBE 
\be
S_{12}(p_{1}, p_{2})\, R_{1}^{R}(p_{1})\, S_{21}(p_{2}, -p_{1})\, 
R_{2}^{R}(p_{2}) = 
R_{2}^{R}(p_{2})\, S_{12}(p_{1}, -p_{2})\, R_{1}^{R}(p_{1})\, 
S_{21}(-p_{2}, -p_{1}) \,.
\label{BYBERY}
\ee 
The algebra (\ref{boundarySRY}) also implies the right boundary unitarity 
equation
\be
R^{R}(p)\, R^{R}(-p) = \id \,.
\label{boundunitarityR}
\ee 
We also assume, in analogy with the bulk case
(\ref{bulkphysicalunitarity}), the physical unitarity condition
\be
R^{R}(p)\, R^{R}(p)^{\dagger} = \id \,.
\label{boundphysicalunitarityR}
\ee 

For a left boundary,  we introduce a 
left boundary operator $B_{L}$, and use the conjugate ZF operators 
$A^{i}(p)$ to define a left boundary $S$-matrix $R^{L}(p)$, 
\footnote{One could try to instead use $A_{i}^{\dagger}(p)$ to 
define a left boundary $S$-matrix, namely
$B_{L}\, A^{\dagger}_{i}(p) = R_{i}^{L\, i'}(p)\, B_{L} \, 
A^{\dagger}_{i'}(-p)$, which would instead obey (cf. (\ref{BYBELYn}))
\be
R_{1}^{L}(p_{1})\, S_{12}(-p_{1}, p_{2})\, R_{2}^{L}(p_{2})\, S_{21}(-p_{2}, 
-p_{1}) = S_{12}(p_{1}, p_{2})\, R_{2}^{L}(p_{2})\,  S_{21}(-p_{2}, 
p_{1})\, R_{1}^{L}(p_{1}) \,.
\label{BYBELY}
\ee
However, this left boundary $S$-matrix would {\it not} obey the 
natural relation (\ref{leftright}).}
\be
B_{L}\, A^{i}(p) = R^{L\, i}_{\, i'}(p)\, B_{L} \, 
A^{i'}(-p) \,.
\label{boundarySLY}
\ee
If we identify $B_{L}$ with $\left( B_{R} \right)^{\dagger}$, then 
(\ref{boundarySRY}) and (\ref{boundarySLY}) imply 
\be
R^{L}(p) = R^{R}(p)^{\dagger} \,.
\ee
Hence, it suffices to consider only the case of right boundary
scattering.  The unitarity conditions (\ref{boundunitarityR}),
(\ref{boundphysicalunitarityR}) then imply the relation
\be
R^{L}(p) = R^{R}(-p) 
\label{leftright}
\ee
which was proposed by HM. We remark that, starting from 
$B_{L}\, A^{i}(p)\,  A^{j}(p)$, 
and with the help of (\ref{bulkSconjugate}), one can derive the left BYBE
\be
\lefteqn{R_{1}^{L\, t_{1}}(p_{1})\, S_{12}^{t_{1}t_{2}}(-p_{1}, p_{2})\, 
R_{2}^{L\, t_{2}}(p_{2})\, S_{21}^{t_{1}t_{2}}(-p_{2}, 
-p_{1})} \non \\
& & = S_{12}^{t_{1}t_{2}}(p_{1}, p_{2})\, R_{2}^{L\, 
t_{2}}(p_{2})\,  S_{21}^{t_{1}t_{2}}(-p_{2}, 
p_{1})\, R_{1}^{L\, t_{1}}(p_{1}) \,,
\label{BYBELYn}
\ee
where $t_{i}$ denotes transposition in the $i^{th}$ space.
Taking the transpose in both spaces 1 and 2, interchanging spaces 1 
and 2 (i.e., conjugating both sides with the permutation matrix 
${\cal P}_{12}$), and relabeling $p_{2}\mapsto -p_{1}\,,  
p_{1}\mapsto -p_{2}$,  we recover the right BYBE (\ref{BYBERY}) with 
the identification  (\ref{leftright}).

Following HM, we proceed to determine the boundary $S$-matrix using
the symmetry of the problem.  The $Y=0$ giant graviton brane preserves
only an $su(1|2)$ subalgebra \cite{HM}, which includes (say) the 
supersymmetry generators $Q_{\alpha}^{\ 1}$ and $Q_{1}^{\dagger 
\alpha}$ with $\alpha \in \{3\,, 4\}$. The right boundary $S$-matrix is 
diagonal, with matrix elements 
\be 
R_{\, 1}^{R\, 1} = r_{1}\,, \quad R_{\, 2}^{R\, 2} = r_{2}\,, 
\quad R_{\, 3}^{R\, 3} =  R_{\, 4}^{R\, 4} = r \,. 
\ee
Using first  (\ref{repBulk2}) and then (\ref{boundarySRY}), we find
\be
Q_{3}^{\ 1}\,  A^{\dagger}_{1}(p)\, B_{R} |0\rangle
= e^{-i p/2}\left[ a(p)  A^{\dagger}_{3}(p) +  A^{\dagger}_{1}(p)\, 
Q_{3}^{\ 1} \right] B_{R} |0\rangle = e^{-i p/2} a(p) r  
A^{\dagger}_{3}(-p) B_{R} |0\rangle\,,
\label{result1}
\ee
where we have passed to the second equality using also the
assumption that $Q_{3}^{\ 1}$ annihilates the vacuum state
$B_{R} |0\rangle$.  Reversing the order, i.e., using first
(\ref{boundarySRY}) and then (\ref{repBulk2}), we obtain
\be
Q_{3}^{\ 1}\,  A^{\dagger}_{1}(p)\, B_{R} |0\rangle
&=& r_{1} Q_{3}^{\ 1}\,  A^{\dagger}_{1}(-p)\, B_{R} |0\rangle
= r_{1} e^{i p/2}\left[ a(-p)  A^{\dagger}_{3}(-p) +  A^{\dagger}_{1}(-p)\, 
Q_{3}^{\ 1} \right] B_{R} |0\rangle \non \\
&=&  r_{1} e^{i p/2} a(-p) A^{\dagger}_{3}(-p) B_{R} |0\rangle\,.
\label{result2}
\ee
Consistency of the results (\ref{result1}) and (\ref{result2}) requires
\be
\frac{r_{1}}{r}=e^{-i p} \frac{a(p)}{a(-p)} =e^{-i p}\,,
\ee
where, in passing to the second equality, we have used \cite{HM}
\be
x^{\pm}(-p) = - x^{\mp}(p) \,, \qquad \eta(-p) = \eta(p) \,,
\label{negation}
\ee 
since $x^{\pm} \mapsto -x^{\mp}$ corresponds to $p \mapsto -p\,, \ H \mapsto H$.
Similarly, starting from $Q_{3}^{\ 1}\,  A^{\dagger}_{4}(p)\, B_{R} 
|0\rangle$, we readily obtain
\be
\frac{r_{2}}{r}=e^{i p} \frac{b(-p)}{b(p)} =-1\,.
\ee
The same results are obtained using instead the other conserved 
supersymmetry generators.
We conclude that the right boundary $S$-matrix is given by the 
diagonal matrix \footnote{The left boundary $S$-matrix (\ref{boundarySLY}) can be computed in a 
completely analogous manner using the Hermitian conjugate of the 
relations (\ref{repBulk2}) with $\left( Q_{\alpha}^{\ 
a}\right)^{\dagger} = Q_{a}^{\dagger \alpha}$. The result is an 
accord with (\ref{leftright}).}
\be
R^{R}(p) = R^{R}_{0}(p) \diag( e^{-ip}\,, -1 \,, 1 \,, 1 ) \,.
\label{boundarySRY2}
\ee
We have explicitly verified that this matrix satisfies the 
standard BYBE (\ref{BYBERY}). It also 
evidently satisfies the boundary unitarity equation (\ref{boundunitarityR}), 
provided that the corresponding scalar factor satisfies
\be
R_{0}^{R}(p)\, R_{0}^{R}(-p) = 1 \,.
\label{boundaryunitarity2}
\ee 
If we demand the conservation of the supersymmetry 
generators $Q_{\alpha}^{\ 2}\,, Q_{2}^{\dagger \alpha}$ instead of
$Q_{\alpha}^{\ 1}\,, Q_{1}^{\dagger \alpha}$, then we obtain the 
same result (\ref{boundarySRY2}) except with the first two elements permuted.

The matrix (\ref{boundarySRY2}) is similar (but not identical) to 
the right boundary $S$-matrix proposed by HM.  The latter does not satisfy 
(\ref{BYBERY}), but it does satisfy (\ref{BYBELY}). We note that 
the left HM boundary $S$-matrix and our right boundary $S$-matrix are 
related by
\be
R^{L}(p)_{HM} = R^{R}(p)\, U(2p)
\label{relationtoHMY}
\ee
(up to a permutation of the first two elements),
where $U(p)$ is a diagonal matrix relating the ``string'' and ``chain'' bases
given by (see Eq. (8.8) in \cite{AFZ})
\be
U(p) = \diag( e^{ip/2}\,, e^{ip/2}\,, 1 \,, 1 ) \,.
\label{Umatrix}
\ee 

One can show that the boundary $S$-matrix (\ref{boundarySRY2}) is
essentially (i.e., up to permutations, etc.)  the unique diagonal
solution of the BYBE (\ref{BYBERY}) with the AFZ bulk $S$-matrix.  In
particular, no free boundary parameters appear in the solution.  This
is different from the case of the Hubbard model \cite{Sh}, for which
the BYBE has diagonal solutions with a free parameter \cite{hubbard}.
This difference seems paradoxical, given that the AFZ $S$-matrix is
related \cite{MM} to Shastry's $R$-matrix.  This difference can be
attributed to the fact that a specific parametrization of $x^{\pm}(p)$
is needed to relate the bulk matrices (see Eqs.  (12), (14) and (A.3)
in \cite{MM}), which is incompatible with the boundary matrices in
\cite{hubbard}.

\subsection{$Z=0$ giant graviton brane}\label{subsec:Z}

According to HM, the $Z=0$ giant graviton brane has a boundary degree 
of freedom and full $su(2|2)$ symmetry. Correspondingly, we introduce 
a right boundary operator with an index $B_{j\, R}$,
\be
A^{\dagger}_{i}(p)\, B_{j\, R} = R_{\ i\, j}^{R\, i' j'}(p)\, A^{\dagger}_{i'}(-p)\,  
B_{j'\, R} \,,
\label{boundarySRZ}
\ee
and we arrange the boundary $S$-matrix elements into the $16 \times 
16$ matrix $R^{R}$,
\be
R^{R} = R_{\ i\, j}^{R\,i' j'} e_{i\, i'}\otimes e_{j\, j'}\,.
\label{boundSRZ2}
\ee
It satisfies the right BYBE (cf. Eq. (\ref{BYBERY}))
\be
S_{12}(p_{1}, p_{2})\, R_{13}^{R}(p_{1})\, S_{21}(p_{2}, -p_{1})\, 
R_{23}^{R}(p_{2}) = 
R_{23}^{R}(p_{2})\, S_{12}(p_{1}, -p_{2})\, R_{13}^{R}(p_{1})\, 
S_{21}(-p_{2}, -p_{1}) \,,
\label{BYBERZ}
\ee 
and the right boundary unitarity equation (\ref{boundunitarityR}),
where now $\id$ is the 16-dimensional identity matrix.

Moreover, we introduce the left boundary operator $B_{L}^{\, i} =
\left( B_{i\, R} \right)^{\dagger}$, and define the left boundary
$S$-matrix by
\be
B_{L}^{\, i}\, A^{j}(p) = R^{L\, i\, j}_{\, i' j'}(p)\, B_{L}^{\, i'} \, 
A^{j'}(-p) \,.
\label{boundarySLZ}
\ee
It follows from (\ref{boundarySRZ}) and (\ref{boundarySLZ}) that
\be
R_{12}^{L}(p) = R_{21}^{R}(p)^{t_{1}t_{2}\, *} \equiv 
R_{21}^{R}(p)^{\dagger}\,.
\ee 
The unitarity conditions (\ref{boundunitarityR}), (\ref{boundphysicalunitarityR})
then imply a relation analogous to the one for the $Y=0$ case 
(\ref{leftright}),
\be
R_{12}^{L}(p) = R_{21}^{R}(-p)\,.
\label{leftrightZ}
\ee 

We again use symmetry to compute the boundary $S$-matrix. We assume
that the symmetry generators act on the right boundary operators as 
follows 
\be
L_{a}^{\ b}\, B_{c\, R} &=& (\delta_{c}^{b}\delta_{a}^{d} - 
\frac{1}{2}\delta_{a}^{b}\delta_{c}^{d}) B_{d\, R} \,, \quad 
L_{a}^{\ b}\, B_{\gamma\, R} = 0 \,, \non \\
R_{\alpha}^{\ \beta}\, B_{\gamma\, R} &=& 
(\delta_{\gamma}^{\beta}\delta_{\alpha}^{\delta} - 
\frac{1}{2}\delta_{\alpha}^{\beta}\delta_{\gamma}^{\delta}) 
B_{\delta\, R} \,, \quad
R_{\alpha}^{\ \beta}\, B_{c\, R} =  0\,,
\label{repBoundaryR1}
\ee
and \footnote{If we had used the commutation relations of the ZF 
operators with the supersymmetry generators preferred 
by AFZ (namely, Eq. (4.15) in \cite{AFZ}) instead of (\ref{repBulk2}), then 
operators $e^{\pm i P/2}$ would appear on the RHS of (\ref{repBoundaryR2}).}
\be
Q_{\alpha}^{\ a}\, B_{b\, R} &=&
a_{B} \delta_{b}^{a} B_{\alpha\, R} \,, \non \\
Q_{\alpha}^{\ a}\, B_{\beta\, R} &=&
b_{B} \epsilon_{\alpha \beta}\epsilon^{a b} B_{b\, R}\,, \non \\
Q_{a}^{\dagger \alpha}\, B_{b\, R} &=& 
c_{B} \epsilon_{a b} \epsilon^{\alpha \beta} B_{\beta\, R} \,, \non \\
Q_{a}^{\dagger \alpha}\, B_{\beta\, R} &=&
d_{B} \delta_{\beta}^{\alpha} B_{a\, R} \,.
\label{repBoundaryR2}
\ee 
The boundary operators form a fundamental representation of 
the symmetry algebra (\ref{symmetryalgebra}) provided
\be
a_{B} d_{B} - b_{B} c_{B} = 1 \,, 
\label{short}
\ee
with 
\be
C=a_{B} b_{B}\,, \quad  C^{*}=c_{B} d_{B}\,, \quad H = a_{B} d_{B} + 
b_{B} c_{B} \,.
\ee
We take $d_{B}= a_{B}^{*}\,, c_{B} =  b_{B}^{*}$ (unitarity); and we 
set $C = ig$, which is consistent with the requirement $|C|=g$  \cite{HM}.
A suitable parametrization is
\be
a_{B} = \sqrt{g}\eta_{B}\,, \quad 
b_{B} = \sqrt{g}\frac{i}{\eta_{B}}\,, \quad 
c_{B} = \sqrt{g}\frac{\eta_{B}}{x_{B}}\,, \quad 
d_{B} =\sqrt{g}\frac{x_{B}}{i \eta_{B}}\,,
\label{BoundaryParameters}
\ee
where
\be
\eta_{B} = \sqrt{-i x_{B}}\,, \qquad 
x_{B}=\frac{i}{2g}\left(1+\sqrt{1+4g^{2}}\right) \,.
\label{BoundaryParameters2}
\ee
This parametrization coincides with the one used by HM for a
particular value of their parameter $f_{B}$, namely $f_{B}=i$.  (See
Eqs.  (3.34) - (3.37) in \cite{HM}.)  We emphasize that our parameters
(\ref{BoundaryParameters}) are independent of $p$, in keeping with 
the fact that momentum is a property only of the bulk excitations.
In contrast, because HM use Beisert's ``non local'' notation (see the second 
reference in \cite{Be}), their values of $f_{B}$ are functions of $p$  
which change under scattering. 

The nonzero matrix elements of the right boundary $S$-matrix are
\be
R_{\ a\, a}^{R\, a\, a} &=& A\,, \quad 
R_{\ \alpha\, \alpha}^{R\, \alpha\, \alpha} = D\,, \non \\
R_{\ a\, b}^{R\, a\, b} &=& \frac{1}{2}(A+B)\,, \quad 
R_{\ a\, b}^{R\, b\, a} = \frac{1}{2}(A-B) \,, \non \\
R_{\ \alpha\, \beta}^{R\, \alpha\, \beta} &=& \frac{1}{2}(D+E)\,, \quad 
R_{\ \alpha\, \beta}^{R\, \beta\, \alpha} = \frac{1}{2}(D-E) \,, \non \\
R_{\ a\, b}^{R\, \alpha\, \beta} &=& 
\frac{1}{2}\epsilon_{a b}\epsilon^{\alpha \beta}\, C \,, \quad
R_{\ \alpha\, \beta}^{R\, a\, b} = 
\frac{1}{2}\epsilon^{a b}\epsilon_{\alpha \beta}\, F \,, \non \\
R_{\ a\, \alpha}^{R\, a\, \alpha} &=& K\,, \quad 
R_{\ a\, \alpha}^{R\, \alpha\, a} = L \,, \quad 
R_{\ \alpha\, a}^{R\, a\, \alpha} = G\,, \quad 
R_{\ \alpha\, a}^{R\, \alpha\, a} = H \,,  
\label{boundarySRZ3}
\ee
where $a\,, b \in \{1\,, 2\}$ with $a \ne b$; and 
$\alpha\,, \beta \in \{3\,, 4\}$ with $\alpha \ne \beta$. Proceeding 
as before, we obtain
\be
A &=&  e^{-2ip} A_{HM} = 
 R_{0}^{R}\frac{x^{-}(x^{+}+x_{B})}{x^{+}(x^{-}-x_{B})}\,, \non \\
B &=&  e^{-2ip} B_{HM} = 
 R_{0}^{R}\frac{2x^{+}x^{-}x_{B}+(x^{+}-x_{B})[-2(x^{+})^{2}+2(x^{-})^{2}+x^{+}x^{-}]}
{(x^{+})^{2}(x^{-}-x_{B})}
\,, \non \\
C &=&   C_{HM} =
 R_{0}^{R}\frac{2 \eta \eta_{B}}{i}\frac{(x^{-}+x^{+})(x^{-} x_{B}-x^{+} 
x_{B}-x^{-}x^{+})}{x_{B}x^{-}(x^{+})^{2}(x^{-}-x_{B})}\,, \qquad
D = D_{HM} =  R_{0}^{R}\,, \non \\
E &=&   E_{HM} =
 R_{0}^{R}\frac{2[(x^{+})^{2}-(x^{-})^{2}][-x^{+}x^{-}+x_{B}(x^{-}-x^{+}+x^{-}(x^{+})^{2}]
-x_{B}(x^{+}x^{-})^{2}(x_{B}-x^{-})}{(x^{-}x^{+})^{2}x_{B}(x^{-}-x_{B})}
\,, \non \\
F &=&  e^{-2ip} F_{HM} = 
 R_{0}^{R}\frac{2i}{\eta \eta_{B}} 
\frac{[(x^{+})^{2}-(x^{-})^{2}](x_{B}x^{+}-x_{B}x^{-}+x^{+}x^{-})}{(x^{+})^{2}x^{-}(x^{-}-x_{B})}
\,, \non \\
G &=& \ e^{-ip} G_{HM} =
 R_{0}^{R}\frac{\eta_{B}}{\eta}\frac{(x^{+})^{2}-(x^{-})^{2}}{x^{+}(x^{-}-x_{B})}\,, \qquad 
H = e^{-ip} H_{HM} =
 R_{0}^{R}\frac{(x^{+})^{2}-x_{B} x^{-}}{x^{+}(x^{-}-x_{B})}\,, \non \\
K &=& e^{-ip} K_{HM} =
 R_{0}^{R}\frac{(x^{-})^{2}+x_{B} x^{+}}{x^{+}(x^{-}-x_{B})}\,, \qquad 
L = e^{-ip} L_{HM} = 
 R_{0}^{R}\frac{\eta}{\eta_{B}}\frac{(x^{+} + x^{-})x_{B}}{x^{+}(x^{-}-x_{B})}\,, 
\label{boundarySRZ4}
\ee
where $A_{HM}$, etc.  are the corresponding HM amplitudes for the {\it
left} boundary $S$-matrix (see Eq.  (3.46) in \cite{HM}) with $f=i$.
We have explicitly verified that the right BYBE (\ref{BYBERZ}) is
satisfied, as well as the boundary unitarity equation
(\ref{boundunitarityR}), provided that the scalar factor obey
(\ref{boundaryunitarity2}).

We note that the left HM boundary $S$-matrix and our right boundary
$S$-matrix are related by (cf. Eq. (\ref{relationtoHMY}))
\be
R^{L}(p)_{HM} = R^{R}(p)\, U(2p) \otimes U(2p) \,,
\label{relationtoHMZ}
\ee
where $U(p)$ is given by (\ref{Umatrix}).

\section{Crossing relations and scalar factors}\label{sec:crossing}

We turn now to the derivation of crossing relations, which (together
with the unitarity relations) help determine the scalar factors of the
$S$-matrices.  For the boundary $S$-matrices, the crossing relations
and scalar factors are similar to (but not the same as) those for the
HM boundary $S$-matrices.

\subsection{Bulk}

For the bulk $S$-matrix, a crossing relation was first proposed by
Janik \cite{Ja} based on a Hopf algebra structure of the symmetry
algebra.  AFZ subsequently gave an alternative derivation of the
crossing relation based on the ZF algebra.  We now reformulate in
terms of ZF operators yet another derivation of the crossing relation,
due to Beisert \cite{Be}, which is particularly convenient to
generalize to the boundary case \cite{HM}.  To this end, we define the
``singlet'' operator
\be
I(p) = C^{i j}(p)\, A^{\dagger}_{i}(p)\, A^{\dagger}_{j}(\bar 
p) \equiv 
\cf(p)\, \epsilon^{a b} A^{\dagger}_{a}(p)\, A^{\dagger}_{b}(\bar 
p) + \epsilon^{\alpha \beta} A^{\dagger}_{\alpha}(p)\, 
A^{\dagger}_{\beta}(\bar p) \,,
\label{singlet}
\ee
where (as before) $a\,, b \in \{1\,, 2\}$, $\alpha\,, \beta \in 
\{3\,, 4\}$, and the function $\cf(p)$ is yet to be determined. 
Hence, $C(p)$ is the $4 \times 4$ matrix
\be
C(p)= \left( \begin{array}{cccc}
0  &\cf(p)  & 0 & 0 \\
-\cf(p) &0  &0 & 0 \\
0  &0  &0 & 1 \\
0  &0  &-1 & 0
\end{array} \right) \,.
\label{CCmatrix}
\ee 
Moreover, $\bar p$ denotes the antiparticle momentum, with \cite{Ja, AFZ} 
\be
x^{\pm}(\bar p) = \frac{1}{x^{\pm}(p)} \,,
\label{barp}
\ee
since $x^{\pm} \mapsto 1/x^{\pm}$ corresponds to $p \mapsto -p \equiv \bar 
p\,, \ H \mapsto -H \equiv \bar H$.
One can readily check (with the help of Eq.  (\ref{repBulk1})) that
the singlet operator commutes with the bosonic generators.
The function $\cf(p)$ is determined by the condition that the singlet 
operator also commute with the supersymmetry generators. Indeed,
the condition $Q_{3}^{\ 1} I(p) |0\rangle = I(p)\, Q_{3}^{\ 1}  
|0\rangle =0$ readily leads (with the help of Eq. (\ref{repBulk2})) to
\be
\cf(p) = e^{i p/2} \frac{b(\bar p)}{a(p)} = -e^{-i p/2} \frac{b(p)}{a(\bar 
p)} = -i \sgn(p) \,.
\ee
This computation evidently parallels the one in AFZ for the charge 
conjugation matrix. However, the matrix (6.8) in \cite{AFZ} is proportional to 
our $C(-p)$. \footnote{In fact, the momentum dependence of the charge 
conjugation matrix is spurious and can be removed by properly resolving the 
branch cut ambiguity as noticed in \cite{AF2}.}

The crossing relation follows from the requirement that the singlet 
operator scatter trivially with a particle. Indeed,
\be
A^{\dagger}_{i}(p_{1})\, I(p_{2}) &=&
C^{j k}(p_{2})\, A^{\dagger}_{i}(p_{1})\, A^{\dagger}_{j}(p_{2})\, 
A^{\dagger}_{k}(\bar p_{2})\non\\
&=& C^{j k}(p_{2})\, S_{i j}^{i' j'}(p_{1},p_{2})\, A^{\dagger}_{j'}(p_{2})\,
A^{\dagger}_{i'}(p_{1})\, A^{\dagger}_{k}(\bar p_{2})\non\\
&=& C^{j k}(p_{2})\, S_{i j}^{i' j'}(p_{1},p_{2})\, S_{i' k}^{i'' k'}(p_{1}, \bar p_{2})\, 
A^{\dagger}_{j'}(p_{2})\, A^{\dagger}_{k'}(\bar p_{2})\, A^{\dagger}_{i''}(p_{1}) \non\\
&\equiv& I(p_{2})\, A^{\dagger}_{i}(p_{1})
\ee
implies the relation
\be
C^{j k}(p_{2})\, S_{i j}^{i' j'}(p_{1},p_{2})\, S_{i' k}^{i'' k'}(p_{1}, \bar p_{2}) =
C^{j' k'}(p_{2})\, \delta_{i}^{i''} \,,
\ee
which can be re-expressed in matrix notation as
\be
S_{12}^{t_{2}}(p_{1},p_{2})\, C_{2}(p_{2})\, S_{12}(p_{1}, \bar p_{2})\, 
C_{2}(p_{2})^{-1}  =  \id \,.
\ee 
Substituting the result (\ref{bulkS3}), (\ref{bulkS4}) for the 
$S$-matrix, we obtain a crossing relation for the bulk scalar 
factor
\be
S_{0}(p_{1},p_{2})\, S_{0}(p_{1}, \bar p_{2}) = \frac{1}{f(p_{1},p_{2})} \,,
\label{bulkcrossing1}
\ee
where \cite{Ja}
\be
f(p_{1},p_{2}) = \frac{\left(\frac{1}{x^{+}_{1}} - 
x^{-}_{2}\right)(x^{+}_{1} - x^{+}_{2})}
{\left(\frac{1}{x^{-}_{1}} - 
x^{-}_{2}\right)(x^{-}_{1} - x^{+}_{2})} \,.
\label{ffunc}
\ee
Similarly, by demanding $I(\bar p_{1})\, A^{\dagger}_{k}(p_{2}) 
= A^{\dagger}_{k}(p_{2})\, I(\bar p_{1})$ and using the fact that the 
matrix $C(p)$ is antisymmetric, one can also formally obtain 
\be
S_{12}^{t_{1}}(p_{1},p_{2})\, C_{1}(\bar p_{1})\, S_{12}(\bar p_{1}, p_{2})\, 
C_{1}(\bar p_{1})^{-1}  =  \id \,,
\ee 
which implies a second crossing relation for the bulk scalar 
factor  \cite{AFZ}
\be
S_{0}(p_{1}, p_{2})\, S_{0}(\bar p_{1},p_{2}) = \frac{1}{f(p_{1},p_{2})} \,.
\label{bulkcrossing2}
\ee

The crossing equations (\ref{bulkcrossing1}), (\ref{bulkcrossing2}) 
corresponding to the AFZ (string) $S$-matrix
are the same as Janik's relations \cite{Ja} corresponding to 
Beisert's (spin chain) $S$-matrix \cite{Be}, except the right-hand-sides are inverted. 
Correspondingly, the solutions are also inversely related.

In more detail, let us now now consider the full 
theory, for which there are two $su(2|2)$ factors. Setting \cite{AFS, AF} 
\footnote{For the spin chain $S$-matrix, the RHS of (\ref{AFrelation}) is 
inverted \cite{HM, BHL}.}
\be
S_{0}(p_{1}\,, p_{2})^{2} = \frac{x^{-}_{1}-x^{+}_{2}}{x^{+}_{1}-x^{-}_{2}}
\frac{1-\frac{1}{x^{+}_{1}x^{-}_{2}}}{1-\frac{1}{x^{-}_{1}x^{+}_{2}}}
\sigma(p_{1}\,, p_{2})^{2} \,,
\label{AFrelation}
\ee
the crossing equations (\ref{bulkcrossing1}), (\ref{bulkcrossing2})
imply that the ``dressing factor'' $\sigma(p_{1}\,, p_{2})$ obeys
\be
\sigma(\bar p_{1}, p_{2})\, \sigma( p_{1},p_{2}) = 
\frac{x^{-}_{2}}{x^{+}_{2}}\frac{1}{f(p_{1},p_{2})} \,, \qquad 
\sigma(p_{1}, \bar p_{2})\, \sigma(p_{1}, p_{2}) = 
\frac{x^{+}_{1}}{x^{-}_{1}}\frac{1}{f(p_{1},p_{2})} \,,
\label{sigmacrossingrelations}
\ee
and the unitarity equation (\ref{bulkunitarityscalar}) implies
\be
\sigma(p_{1}, p_{2})\, \sigma(p_{2}, p_{1}) = 1 \,.
\label{sigmaunitarityrelations}
\ee
The relations (\ref{sigmacrossingrelations}),
(\ref{sigmaunitarityrelations}) are ``universal'' in the sense that
the dressing factor for the spin chain $S$-matrix obeys the same
relations \cite{BHL}.  A solution is given by \cite{AF}-\cite{DHM}
\be
\sigma(x^{\pm}_{1}\,, x^{\pm}_{2}) = 
\frac{R(x^{+}_{1}\,, x^{+}_{2})\ R(x^{-}_{1}\,, x^{-}_{2})}
{R(x^{+}_{1}\,, x^{-}_{2})\ R(x^{-}_{1}\,, x^{+}_{2})} \,,
\qquad
R(x_{1}\,, x_{2}) = e^{i\left[ \chi(x_{1}\,, x_{2}) - \chi(x_{2}\,, 
x_{1}) \right]} \,,
\label{sigma}
\ee
where \cite{DHM}
\be
\chi(x_{1}\,, x_{2}) =-i \oint_{|z_{1}| = 1} \frac{dz_{1}}{2\pi} 
\oint_{|z_{2}| = 1} \frac{dz_{2}}{2\pi} 
\frac{\ln \Gamma\left(1 + i 
g(z_{1}+\frac{1}{z_{1}}-z_{2}-\frac{1}{z_{2}})\right)}
{(x_{1}-z_{1})(x_{2}-z_{2})} \,.
\label{chi}
\ee

\subsection{Boundary: $Y=0$ giant graviton brane}

For the boundary case, we follow HM and consider the scattering of 
the singlet operator (\ref{singlet}) off the boundary. For the right 
boundary, we obtain
\be
I(p)\, B_{R} &=& C^{i j}(p)\, A^{\dagger}_{i}(p)\, A^{\dagger}_{j}(\bar p) 
\, B_{R} \non \\
&=& C^{i j}(p)\, R_{j}^{R\, j'}(\bar p)\,  A^{\dagger}_{i}(p)\, 
A^{\dagger}_{j'}(-\bar p)\, B_{R} \non \\
&=& C^{i j}(p)\, R_{j}^{R\, j'}(\bar p)\,  S_{i j'}^{i' j''}(p,-\bar p)\,
A^{\dagger}_{j''}(-\bar p)\, A^{\dagger}_{i'}(p)\, B_{R} \non \\
&=& C^{i j}(p)\, R_{j}^{R\, j'}(\bar p)\,  S_{i j'}^{i' j''}(p,-\bar p)\,
R_{i'}^{R\, i''}(p)\, 
A^{\dagger}_{j''}(-\bar p)\, A^{\dagger}_{i''}(-p)\, B_{R} \non \\
&\equiv& I(-\bar p)\,  B_{R} \,,
\label{boundcrossing}
\ee
which implies the relation
\be
C^{i j}(p)\, R_{j}^{R\, j'}(\bar p)\,  S_{i j'}^{i' j''}(p,-\bar p)\,
R_{i'}^{R\, i''}(p) = C^{j'' i''}(p) \,.
\ee
Substituting the results for the bulk (\ref{bulkS3}), (\ref{bulkS4})
and boundary (\ref{boundarySRY2}) $S$-matrices, we obtain the right boundary
crossing relation
\be
R^{R}_{0}(p)\, R^{R}_{0}(\bar p)\, S_{0}(p,-\bar p) = 
\frac{1}{h_{b}(-p)}  = h_{b}(p) \,,
\label{bouncrossingY0R}
\ee
where \cite{HM}
\be
h_{b}(p) =\frac{\frac{1}{x^{-}} + x^{-}}{\frac{1}{x^{+}} + x^{+}} \,.
\label{hb}
\ee
The boundary crossing relation (\ref{bouncrossingY0R}) is similar to the one 
found by Ghoshal and Zamolodchikov \cite{GZ} for relativistic 
integrable theories, and is the same as HM (3.29), except with 
$p \mapsto -p$ in the RHS. 

For the full theory, the crossing relation becomes
\be
R^{R}_{0}(p)^{2}\, R^{R}_{0}(\bar p)^{2}  = 
h_{b}(p)^{2}\frac{1}{S_{0}(p,-\bar p)^{2}} = 
h_{b}(p)\frac{1}{\sigma(p,-\bar p)^{2}} \,,
\label{bouncrossingfullY0R}
\ee
where we have used (\ref{AFrelation}).  Since the RHS is the inverse
of HM's relation (3.31), the solution is the inverse of the solution
found by Chen and Correa (see Eq. (27) in \cite{CC})
\be
R^{R}_{0}(p)^{2} = R^{R}_{0}(p)^{-2}_{HM} = F(p)\, \sigma(p \,, -p) \,,
\label{Y0Rresult}
\ee
where we have used (\ref{sigmaunitarityrelations}), and $F(p)$ is a
CDD-type factor obeying
\be
F(p)\, F(\bar p) = 1\,, \qquad F(p)\, F(-p) = 1 \,.
\label{CDD}
\ee 

\subsection{Boundary: $Z=0$ giant graviton brane}

For the right $Z=0$ boundary, a calculation analogous to (\ref{boundcrossing})
implies the relation
\be
C^{i j}(p)\, R_{j k}^{R\, j' k'}(\bar p)\,  S_{i j'}^{i' j''}(p,-\bar p)\,
R_{i' k'}^{R\, i'' k''}(p) = C^{j'' i''}(p)\, \delta_{k}^{k''}\,.
\ee
Substituting the results for the bulk (\ref{bulkS3}), (\ref{bulkS4})
and boundary (\ref{boundarySRY2}) $S$-matrices, we obtain the right
boundary crossing relation
\be
R^{R}_{0}(p)\, R^{R}_{0}(\bar p)\, S_{0}(p,-\bar p) = 
\frac{1}{h_{b}(-p) h_{B}(-p)} = \frac{h_{b}(p)}{h_{B}(-p)} \,,
\label{bouncrossingZ0R}
\ee
where \cite{CC, ABR}
\be
h_{B}(p) &=& 
\frac{x^{+}}{x^{-}}\left(\frac{x_{B}-x^{-}}{x_{B}-x^{+}}\right)
\frac{1+(x_{B} x^{-}x^{+})^{2}}{(1-(x_{B}x^{+})^{2})(1-x^{-}x^{+})}\non \\
&=& \left(\frac{x_{B}-x^{-}}{x_{B}-x^{+}}\right)
\left( \frac{\frac{1}{x^{-}}+x_{B}}{\frac{1}{x^{+}}+x_{B}}\right) \,.
\label{hB}
\ee
The boundary crossing relation (\ref{bouncrossingZ0R}) is the same as the one
found in \cite{CC}, except with $p \mapsto -p$ in the RHS. 

For the full theory, the crossing relation becomes 
\be
R^{R}_{0}(p)^{2}\, R^{R}_{0}(\bar p)^{2}  = 
\frac{h_{b}(p)^{2}}{h_{B}(-p)^{2}} \frac{1}{S_{0}(p,-\bar p)^{2}} =
\frac{h_{b}(p)}{h_{B}(-p)^{2}} \frac{1}{\sigma(p,-\bar p)^{2}}\,.
\ee
Comparing with the corresponding $Y=0$ results (\ref{bouncrossingfullY0R}), 
(\ref{Y0Rresult}), we see that
\be
R^{R}_{0}(p)^{2} = F(p)\, \sigma(p \,, -p)\, \tilde R^{R}_{0}(p)^{2} \,,
\label{Z0Rresult}
\ee
where
\be
\tilde R^{R}_{0}(p)^{2}\, \tilde R^{R}_{0}(\bar p)^{2} = 
\frac{1}{h_{B}(-p)^{2}} \,, \qquad 
\tilde R^{R}_{0}(p)^{2}\, \tilde R^{R}_{0}( -p)^{2} = 1 \,.
\ee
We solve for $\tilde R^{R}_{0}(p)^{2}$ following \cite{ABR}
using the identities
\be
\sigma(p\,, -x_{B})^{2}\, \sigma(\bar p\,, -x_{B})^{2} = 
\frac{h_{b}(p)^{2}}{h_{B}(-p)^{2}} \,, \qquad
\sigma(p\,, -x_{B})^{2}\, \sigma(-p\,, -x_{B})^{2} = 1\,,
\label{ident2}
\ee
which we prove in Appendix \ref{sec:ident2}. We conclude that
\be
\tilde R^{R}_{0}(p)^{2} = \frac{1}{h_{b}(p)} \sigma(p\,, -x_{B})^{2} 
\label{tildeR0Rresult} \,.
\ee 
As noted by HM, the boundary $S$-matrix for the full theory has a
double pole at $x^{-}=x_{B}$ (see Eq. (\ref{boundarySRZ4}) above). It 
can be reduced to a simple pole (corresponding to the second boundary 
bound state \cite{HM}) by choosing the CDD factor
\be
F(p) = \left(\frac{x^{-}-x_{B}}{\frac{1}{x^{-}}-x_{B}}\right)
\left( \frac{\frac{1}{x^{+}}+x_{B}}{x^{+}+x_{B}}\right) \,,
\label{Z0CDD}
\ee
which contains the factor $(x^{-}-x_{B})$ and satisfies (\ref{CDD}).
Summarizing, the right boundary scalar factor
$R^{R}_{0}(p)^{2}$ is given by (\ref{Z0Rresult}),  
(\ref{tildeR0Rresult}) and (\ref{Z0CDD}).

\section{Discussion}\label{sec:conclude}

We have seen that not only bulk \cite{AFZ} but also boundary
$S$-matrices of string/gauge theory can satisfy the usual Yang-Baxter
equation.  The latter are closely related to the boundary $S$-matrices
which were proposed in \cite{HM}, as can be seen from Eqs.
(\ref{relationtoHMY}) and (\ref{relationtoHMZ}).  Presumably, as in
the bulk case, the differences are due to working in different bases.
It should now be possible to bring the well-developed techniques of
the Quantum Inverse Scattering Method to bear on boundary problems in
string/gauge theory.  For example, one can now try to construct the
commuting ``double-row'' transfer matrix \cite{Sk} and determine its
eigenvalues in terms of roots of corresponding Bethe Ansatz equations.
We hope to be able to address these and related problems in the near
future.

\section*{Acknowledgments}
This work was initiated at the 2007 APCTP Focus Program ``Liouville, 
Integrability and Branes (4)''. We thank the participants, and also 
A. Belitsky, for discussions. We are also grateful to G. Arutyunov 
and D. Hofman for reading and commenting on a draft.
This work was supported in part by KRF-2007-313-C00150 (CA) and by the
National Science Foundation under Grants PHY-0244261 and PHY-0554821
(RN).

\begin{appendix}

\section{Derivation of (\ref{ident2})}\label{sec:ident2}

In order to derive the first identity in (\ref{ident2}), we first derive
the more general result \footnote{We denote the momentum dependence 
of functions by $x\,, x^{\pm}\,, p$ (or  $y\,, y^{\pm}\,, p$, etc.) 
interchangeably.}
\be
\sigma( y\,, x_{(n)})^{2}\, \sigma(\bar y\,, x_{(n)})^{2} = 
\left( \frac{x_{(n)}^{-}}{x_{(n)}^{+}}\right)^{2} 
\frac{h( y\,, x_{(n)})^{2}}{f( y\,, x_{(n)})^{2}} \,,
\label{genident2}
\ee
where (cf. (\ref{ffunc}))
\be
f( y\,, x_{(n)}) =
\frac{\left(\frac{1}{y^{+}} - 
x^{-}_{(n)}\right)(y^{+} - x^{+}_{(n)})}
{\left(\frac{1}{y^{-}} - 
x^{-}_{(n)}\right)(y^{-} - x^{+}_{(n)})}  \,, \qquad
h( y\,, x_{(n)}) =
\frac{y^{+}+\frac{1}{y^{+}} -x^{+}_{(n)}-\frac{1}{x^{+}_{(n)}}}
{y^{-}+\frac{1}{y^{-}} -x^{-}_{(n)}-\frac{1}{x^{-}_{(n)}}} \,.
\label{defh}
\ee
Moreover, $x^{\pm}_{(n)}$ are the parameters corresponding to an $n$-magnon bound 
state of momentum $p$ given by \cite{DHM, CDO}
\be
x^{\pm}_{(n)} = \frac{e^{\pm i p/2}}{4 g \sin(p/2)}\left(n 
+\sqrt{n^{2}+16 g^{2} \sin^{2}(p/2)}\right) \,,
\label{boundparams}
\ee
which obey the constraint
\be
x^{+}_{(n)}+ \frac{1}{x^{+}_{(n)}}-x^{-}_{(n)}-\frac{1}{x^{-}_{(n)}} 
= \frac{i n}{g} \,.
\label{bsconstraint}
\ee
The $n$ magnons have momenta $p_{1}\,, p_{2}\,, \ldots\,, p_{n}$ which 
form a composite (Bethe $n$-string), with
\be
x^{-}_{j} = x^{+}_{j-1} \,, \qquad j = 2\,, \ldots \,, n\,,
\label{nstring}
\ee
where $x^{\pm}_{j} \equiv x^{\pm}(p_{j})$. Indeed, since
\be
x^{+}_{j}+ \frac{1}{x^{+}_{j}}-x^{-}_{j}-\frac{1}{x^{-}_{j}} 
= \frac{i}{g} \,, \qquad j = 1\,, \ldots \,, n\,,
\ee
summing over $j$ yields the constraint (\ref{bsconstraint}), where
\be
x^{+}_{(n)} = x^{+}_{n}\,, \qquad x^{-}_{(n)} = x^{-}_{1} \,.
\ee

With the help of (\ref{sigma}), (\ref{nstring}), we obtain
\be
\prod_{j=1}^{n}\sigma(y\,, x_{j}) =
\prod_{j=1}^{n}
\frac{R(y^{+}\,, x^{+}_{j})\ R(y^{-}\,, x^{-}_{j})}
{R(y^{+}\,, x^{-}_{j})\ R(y^{-}\,, x^{+}_{j})}
= \frac{R(y^{+}\,, x^{+}_{(n)})\ R(y^{-}\,, x^{-}_{(n)})}
{R(y^{+}\,, x^{-}_{(n)})\ R(y^{-}\,, x^{+}_{(n)})}
\equiv \sigma(y\,, x_{(n)}) \,.
\label{defsigmabound}
\ee
The LHS of (\ref{genident2}) is therefore given by
\be
\sigma( y\,, x_{(n)})^{2}\, \sigma(\bar y\,, x_{(n)})^{2} 
&=& \prod_{j=1}^{n} \left[ \sigma(y\,, x_{j}) \sigma(\bar y\,, x_{j}) 
\right]^{2}
= \prod_{j=1}^{n} \left[ 
\frac{x^{-}_{j}}{x^{+}_{j}}\frac{1}{f(y,x_{j})}
\right]^{2} \non \\
&=&  \left( \frac{x^{-}_{(n)}}{x^{+}_{(n)}}\right)^{2} 
\prod_{j=1}^{n} \frac{1}{f(y,x_{j})^{2}}
\label{intermed1}\,,
\ee
where we have used (\ref{sigmacrossingrelations}), as well as the relation 
\be
\prod_{j=1}^{n} \frac{x^{-}_{j}}{x^{+}_{j}}
=  \frac{x^{-}_{(n)}}{x^{+}_{(n)}}
\,,
\ee
which follows from (\ref{nstring}).  In order to evaluate
the remaining product in (\ref{intermed1}), we make use of the
decomposition \cite{BHL}
\be
f(y\,, x)^{2} = \left[\frac{f(y\,, x)}{f(\bar y\,, x)}\right]
\left[ f(y\,, x) f(\bar y\,, x) \right] 
\equiv \alpha(y\,, x) \ \beta(y\,, x) \,.
\label{decomp}
\ee
Recalling the definition (\ref{ffunc}), we obtain
\be
\alpha(y\,, x) &=& \frac{f(y\,, x)}{f(\bar y\,, x)} =
\left(\frac{y^{+}-x^{+}}{y^{+}-x^{-}}\right)
\left(\frac{y^{-}-x^{-}}{y^{-}-x^{+}}\right)
\left(\frac{y^{-}-\frac{1}{x^{+}}}{y^{-}-\frac{1}{x^{-}}}\right)
\left(\frac{y^{+}-\frac{1}{x^{-}}}{y^{+}-\frac{1}{x^{+}}}\right) \,, \non 
\\
\beta(y\,, x) &=& f(y\,, x) f(\bar y\,, x) = 
\frac{u(y) - u(x) + \frac{i}{g}}{u(y) - u(x) - \frac{i}{g}}
\,,
\ee
where $u(x)$ is defined as \cite{BHL}
\be
u(x) = x^{+} + \frac{1}{x^{+}} - \frac{i}{2g} 
= x^{-} + \frac{1}{x^{-}} + \frac{i}{2g} \,.
\ee
Note that 
\be
u(x_{j}) = u(x_{j-1}) +  \frac{i}{g} \,.
\label{uj}
\ee
After some algebra, we obtain
\be
\prod_{j=1}^{n} \alpha(y\,, x_{j}) = 
\left(\frac{y^{+}-x^{+}_{(n)})}{y^{+}-x^{-}_{(n)})}\right)
\left(\frac{y^{-}-x^{-}_{(n)})}{y^{-}-x^{+}_{(n)})}\right)
\left(\frac{y^{-}-\frac{1}{x^{+}_{(n)})}}{y^{-}-\frac{1}{x^{-}_{(n)})}}\right)
\left(\frac{y^{+}-\frac{1}{x^{-}_{(n)})}}{y^{+}-\frac{1}{x^{+}_{(n)})}}\right)\,;
\label{prodalpha}
\ee
and, using (\ref{uj}), 
\be
\prod_{j=1}^{n} \beta(y\,, x_{j}) &=&
\left( \frac{u(y) - u(x_{1}) + \frac{i}{g}}{u(y) - u(x_{n}) - \frac{i}{g}} 
\right)
\left( \frac{u(y) - u(x_{1})}{u(y) - u(x_{n})}\right) \non \\
&=& \left( \frac{y^{+} - x^{-}_{(n)}}{y^{-} - x^{+}_{(n)}}\right)
\left( \frac{1 - \frac{1}{y^{+} x^{-}_{(n)}}}
{1 - \frac{1}{y^{-} x^{+}_{(n)}}}\right) \frac{1}{h( y\,, x_{(n)})} \,,
\label{prodbeta}
\ee
where $h( y\,, x_{(n)})$ is defined in (\ref{defh}). Combining the 
results (\ref{decomp}), (\ref{prodalpha}), (\ref{prodbeta}), we eventually obtain
\be
\prod_{j=1}^{n} f(y,x_{j})^{2} = 
\prod_{j=1}^{n} \alpha(y\,, x_{j}) \beta(y\,, x_{j}) 
= \frac{f( y\,, x_{(n)})^{2}}{h( y\,, x_{(n)})^{2}} \,,
\ee
where $f( y\,, x_{(n)})$ is defined in (\ref{defh}). Substituting 
this result into (\ref{intermed1}), we arrive at the desired result 
(\ref{genident2}).

We are finally in a position to prove the first identity in
(\ref{ident2}).  The key point \cite{ABR} is that the boundary bound
state can be regarded as an $n=2$ magnon bound state with momentum $p
= \pi$,
\be
\pm x_{B} = x^{\pm}_{(2)}(p = \pi) \,,
\label{key}
\ee
as follows from (\ref{boundparams}) and the expression 
(\ref{BoundaryParameters2}) for $x_{B}$. 
It follows from (\ref{genident2}) that
\be
\sigma( y\,, x_{B})^{2}\, \sigma(\bar y\,, x_{B})^{2} = 
\frac{h_{b}(y)^{2}}{f( y\,, x_{B})^{2}} \,,
\label{genident2B}
\ee
where $\sigma( y\,, x_{B}) \equiv \sigma( y\,,  x_{(2)}(p = \pi))$
(see Eq. (\ref{defsigmabound})). Moreover, recalling (\ref{defh}),
\be
f(y\,, x_{B}) \equiv f( y\,, x_{(2)}(p = \pi)) =
\frac{\left(\frac{1}{y^{+}} + x_{B}\right)(y^{+} - x_{B})}
{\left(\frac{1}{y^{-}} + x_{B}\right)(y^{-} - x_{B})}    \,,
\label{fBdef}
\ee
and, since $x_{B} + 1/x_{B} = i/g$, 
\be
h( y\,, x_{(2)}(p = \pi)) =
\frac{y^{-}+\frac{1}{y^{-}}}
{y^{+}+\frac{1}{y^{+}}} = h_{b}(y)\,,
\ee
where $h_{b}$ is defined in (\ref{hb}).
Finally, performing in (\ref{genident2B}) the continuation $x_{B} \mapsto -x_{B}$, we obtain
\be
\sigma( y\,, -x_{B})^{2}\, \sigma(\bar y\,, -x_{B})^{2} = 
\frac{h_{b}(y)^{2}}{f( y\,, -x_{B})^{2}} = 
\frac{h_{b}(p)^{2}}{h_{B}(-p)^{2}} \,.
\label{genident2B2}
\ee
The second equality follows from $f( y\,, -x_{B}) = h_{B}(-p)$, where
$h_{B}(p)$ is given by (\ref{hB}).  The result (\ref{genident2B2}) is
the first identity in (\ref{ident2}).

The identity
\be
\chi(x_{1}\,, x_{2}) = \chi(-x_{2}\,, -x_{1})
\ee
follows from (\ref{chi}) by replacing $z_{1,2} \mapsto -z_{1,2}$
and interchanging $z_{1} \leftrightarrow z_{2}$. It then follows 
from (\ref{sigma}) that
\be
R(x_{1}\,, x_{2}) = R(-x_{2}\,, -x_{1}) \,.
\label{Rident}
\ee
The second (unitarity) relation in (\ref{ident2}) follows readily 
from (\ref{defsigmabound}), (\ref{key}) and the 
identities (\ref{sigmaunitarityrelations}), (\ref{Rident}).

\end{appendix}

\end{document}